\documentclass[journal]{IEEEtran}
\ifCLASSINFOpdf
  \usepackage[pdftex]{graphicx}
  \DeclareGraphicsExtensions{.pdf,.jpeg,.png}
\else
   \usepackage[dvips]{graphicx}
\fi
\usepackage{float}
\usepackage[tight,footnotesize]{subfigure}

\usepackage{fixltx2e}

\usepackage{stfloats}

\begin{document}
%
\title{VUPIC \\Virtual Machine Usage Based Placement in IaaS Cloud}
%
%
%

\author{Gaurav~Somani, ~\IEEEmembership{member, ~IEEE,}
		Prateek~Khandelwal,
		and Kapil~Phatnani%
		
\IEEEcompsocitemizethanks{
\IEEEcompsocthanksitem Gaurav Somani, Prateek Khandelwal and Kapil Phatnani are with the Department of CSE, The LNM Institute of Technology, Jaipur, India.\protect \\
E-mail: \{gaurav,prateekk.08,kapil.08\}@lnmiit.ac.in
}
\thanks{}}
\maketitle

\begin{abstract}
Efficient resource allocation is one of the critical performance challenges in an Infrastructure as a Service (IaaS) cloud. Virtual machine (VM) placement and migration decision making methods are integral parts of these resource allocation mechanisms. We present a novel virtual machine placement algorithm which takes performance isolation amongst VMs and their continuous resource usage into account while taking placement decisions. Performance isolation is a form of resource contention between virtual machines interested in basic low level hardware resources (CPU, memory, storage, and networks bandwidth). Resource contention amongst multiple co-hosted neighbouring VMs form the basis of the presented novel approach. Experiments are conducted to show the various categories of applications and effect of performance isolation and resource contention amongst them. A per-VM 3-dimensional Resource Utilization Vector (RUV) has been continuously calculated and used for placement decisions while taking conflicting resource interests of VMs into account. Experiments using the novel placement algorithm: VUPIC, show effective improvements in VM performance as well as overall resource utilization of the cloud.
\end{abstract}


%
\IEEEpeerreviewmaketitle

\ifCLASSOPTIONcaptionsoff
  \newpage
\fi

\section{Introduction}
\label{secInt}
%
%

%
%
%
%
\IEEEPARstart{C}{loud computing} has emerged as a phenomenal technology where computing services are provided over the network, with on-demand elastic resources like computing power, memory, storage and network bandwidth. Virtualization provides an easy solution to the objective of cloud, by facilitating creation of Virtual Machines (VMs) on top of the underlying hardware, or, Physical Machines (PMs), resulting into improved  resource utilization and abstraction. Infrastructure as a Service (IaaS) clouds provides raw VMs on which the user can install their own software or applications.

The issues that a cloud service provider must take care of are elasticity, scalability, migration, and performance isolation, out of which,  performance isolation is a very critical challenge. The resource allocation algorithms take resource requirement of a VM into consideration and changes the allocated resources, thus making it an on demand elastic cloud. Virtual machine placement and migration are an integral part of resource allocation in cloud. Changing resource requirements of VMs can be a critical information for VM placement and migration decisions. In a cloud, placement algorithms are responsible for efficient placement of VMs on physical hosts. Initial placement of a VM and subsequent resource usage based placement form a resource allocation procedure in cloud.

In this paper, a novel VM Placement algorithm is presented which takes into account the application specific resource usage made by the VM. The resources are classified into three categories, CPU, network, and disk I/O. Any application would consume these three basic resources. The collective usage of these resources by the applications will result in the total resource usage by the VM. On the basis of usage of these resources, we classify or differentiate various VMs and apply new VM placement algorithm.

The organization of the paper is as follows, section \ref{secRIRC} discusses the problem of performance isolation and its relation with the formulation of VM placement, section \ref{secE1} introduces an initial experiment, section \ref{secPA} introduces the novel algorithm, section \ref{secE2} discusses the experiment to explore the success to the proposed algorithm, and section \ref{secIoA} discusses the current implementation of the proposed algorithm, Section \ref{secRW} discusses the related work and section \ref{secCFW} describes conclusion and future work.

\section{Role of Performance Isolation and Resource Contention in Virtualized Environment}
\label{secRIRC}

An efficient resource allocation method will optimally place VMs on PMs such that the overall resource utilization is maximized \cite{gartner}. It also incorporate ways of introducing efficient and green data centres \cite{averma,anton}, and increase \textit{return on investment (ROI)}. As stated in section \ref{secInt}, performance isolation is very critical issue, and has been studied extensively \cite{gsomani,dgupta}, yet an ideal situation of performance isolation cannot be achieved. With some initial experiments to understand the co-existential or neighbour dependent behaviour of VMs, the authors inferred that the problem of performance isolation can be improved by reducing the resource contention amongst the VMs on same physical host.\\
Performance isolation can be drawn to the lowest level abstraction of shared resources like cpu, memory, network and disk, for example, disk is continuously being used by multiple processes waiting in I/O queue. Thus, I/O scheduler will play a vital role in resource contention, as studied in \cite{dguptaCPUscheduler}. To review the resource contention properties of various applications (in turn,  VMs), experiment 1 has been   conducted and described in next section.

\section{Experiment 1}
\label{secE1}

The objective of experiment 1 is to observe the co-existential behaviour of different VMs, and how their performance is affected when placed with certain kind of VMs. The VMs used in this experiment exploit the individual resources like CPU, Network, and Disk I/O.

\subsection{Experimental Set-up}

\renewcommand{\arraystretch}{1.5}

\begin{table}[!t]
\centering
\scriptsize
\begin{tabular}{|l||c|c|l|}
\hline
\textbf{Name} & \textbf{Processor} & \textbf{Memory} & \textbf{Function}\\ \hline
\texttt{alpha} &  &  & \texttt{Physical}\\ 
&  & & \texttt{Server 1} \\ \cline{4-4}
\texttt{beta} & \texttt{Intel(r)} &  & \texttt{Physical}\\
&  & & \texttt{Server 2} \\ \cline{4-4}

\texttt{gamma} & \texttt{ Core(TM)} & \texttt{4 GB} & \texttt{Storage }\\ 
& \texttt{i5} & & \texttt{Server (NFS)} \\ \cline{4-4}

\texttt{delta} &  &  & \texttt{Physical }\\ 
&  & & \texttt{Server 3} \\ \hline

\end{tabular}
\caption{Physical Server Details in Cloud}
\label{TMachineConf}
\end{table}

The set-up of the experiment 1 is as follows, machines alpha, beta and gamma are the physical hosts for the VMs, and machine gamma acts as Network File Server (NFS), to hold the disk images of the VMs (refer figure \ref{initPlac}). The configuration for the physical machines is provided in table \ref{TMachineConf}. The configuration for the virtual machines is provided in table \ref{TVM1}.

\subsection{Test VMs}
\label{subsecTestvm}
The experiment was conducted by executing the following benchmarks -
\begin{itemize}
\item \textit{CPU intensive :} \texttt{sysbench} \cite{sysbench} was run in CPU test mode for 120 seconds to calculate the prime numbers from 1 to 80,000. The result generated was the number of events of prime calculation that occurred in 120 seconds.

\item \textit{Disk Intensive :} \texttt{sysbench} was run in fileIO mode for 120 seconds to perform random read and write operations on 128 files of 2GB, the result generated here too was the number of events that occurred in 120 seconds.

\item \textit{Network Intensive :} the \texttt{Apache HTTP} server was installed on the VM, serving a static HTML web page, and was benchmarked as follows, using \texttt{ab} \cite{ab}, to serve 100,000 requests, with 10 concurrent request in each iteration. The result generated here was the average time taken per request, in milliseconds.
\end{itemize}

The machines CPU1 and CPU2 are cloned instances of CPU intensive test VMs specified in section \ref{subsecTestvm}, similarly for DISK1 and DISK2, and for WEB1 and WEB2.

\begin{table}[!t]
\centering
\begin{tabular}{|l||l|l|}
\hline
\textbf{Type} & \textbf{Machines} & \textbf{Initial Host} \\ \hline
\texttt{CPU Intensive} & \texttt{CPU1, CPU2} & \texttt{alpha} \\ \hline
\texttt{Disk Intensive} & \texttt{DISK1, DISK2} & \texttt{beta}\\ \hline
\texttt{Network Intensive} & \texttt{WEB1, WEB2} & \texttt{delta} \\ \hline
\end{tabular}
\caption{VM Classification}
\label{TVM1}
\end{table}

\begin{figure}[!t]
\centering
\subfigure[Initial Setup]{
\includegraphics[scale=0.4]{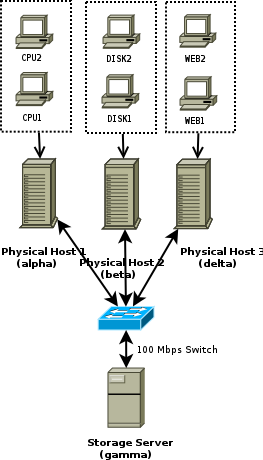}
\label{initPlac}
}
\subfigure[Rearranged Setup]{
\includegraphics[scale=0.4]{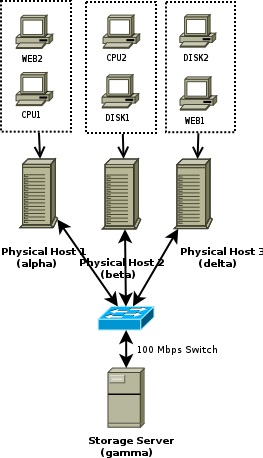}
\label{nuPlac}
}
\label{exp1subfig}
\caption{Setup for Experiment 1}
\end{figure}

The experiment is re-run after rearranging the VMs on physical hosts, providing the following combinations of VMs as CPU intensive with Network intensive, CPU intensive with Disk intensive, Network intensive with Disk intensive (refer figure \ref{nuPlac})

\subsection{Results for the Experiment 1}
The results for experiment 1 are provided in tables \ref{T2.1}, \ref{T2.2}, and \ref{T2.3}. The Tables compares the performances of a particular type of machine, when it is paired with a CPU intensive, Disk Intensive, and Network Intensive machine respectively. The rearrangement as shown in figure \ref{nuPlac} provides for all the possible combinations of VMs and provides a base for comparisons. The results for CPU and Disk Intensive machines were number of events that occurred in the given time, while that for Network Intensive machine was the average time taken for completion of the HTTP requests.

\begin{table}[H]
\scriptsize
\centering
\begin{tabular}{|c|c|c|c|}
\hline
\textbf{-} & \textbf{CPU} & \textbf{Disk} & \textbf{Network} \\ \hline
\textbf{CPU (No. of Events)} & \texttt{5636} & \texttt{7803} & \texttt{5845} \\ \hline
\end{tabular}
\caption{Results of Experiment 1 for CPU Intensive Machine}
\label{T2.1}
\end{table}

\begin{table}[H]
\scriptsize
\centering
\begin{tabular}{|c|c|c|c|}
\hline
\textbf{-} & \textbf{CPU} & \textbf{Disk} & \textbf{Network} \\ \hline
\textbf{Disk (No. of Events)} & \texttt{3900} & \texttt{3000} & \texttt{4400} \\ \hline
\end{tabular}
\caption{Results of Experiment 1 for Disk Intensive Machine}
\label{T2.2}
\end{table}

\begin{table}[H]
\scriptsize
\centering
\begin{tabular}{|c|c|c|c|}
\hline
\textbf{-} & \textbf{CPU} & \textbf{Disk} & \textbf{Network} \\ \hline
\textbf{Network} & \texttt{12.653ms} & \texttt{15.329ms} & \texttt{21.281ms} \\ 
\textbf{(Average Response Time)} & & & \\ \hline
\end{tabular}
\caption{Results of Experiment 1 for Network Intensive Machine}
\label{T2.3}
\end{table}

In table \ref{T2.1}, the increase is performance is significant when kept with a disk intensive VM, which can be attributed to the fact that it will consume a minimal level of CPU time. Similarly, in table \ref{T2.2}, the performance of a disk intensive VM is compromised when kept with another disk intensive VM, and similar is the case for a network intensive VM, as shown in \ref{T2.3}. The results show that for any VM, when kept with another VM that extensively uses the similar kind of resource, the performance drops significantly.

\subsection{Discussion}
\label{SubDiscuss}
The resource usage patterns of each of the VM type can be represented as a three dimensional vector, which will be called as  \textit{Resource Usage Vector (RUV)} in rest of the paper. The components of \textit{RUV} are \textit{cpu, network,} and \textit{disk I/O}, where each component can take value of \textit{L,M,H} and \textit{L} represents \textit{low} usage, \textit{M} represents \textit{medium} usage and \textit{H} represents \textit{High} usage  of the particular component or resource.
\begin{center}
\scriptsize
\begin{equation} \label{eqRUV}
\overrightarrow{RUV}\;=\;\langle\;CPU\;usage,\;Network\;usage,\;Disk\;I/O\;usage\;\rangle
\end{equation}
\end{center}

While running the applications, the resource usage logs were generated for each VM for the entire run of 120 seconds, where the CPU usage, the network usage and the Disk I/O usage for the last 2 seconds per iteration were logged. Using these logs, mean was calculated for each machine for each component (mean is adopted here just for the sake of simplicity, and can be replaced by any other function like maximum magnitude, mean with standard deviation, mean after curve smoothing etc. ). Thus, by setting appropriate range for the mean utilization data, resource vector for the machines came out to be:

\begin{center}
\scriptsize
\textit{CPU Intensive} = \texttt{<H,L,L>}
\\
\textit{Network Intensive} = \texttt{<M,H,L>}
\\
\textit{Disk Intensive} = \texttt{<L,L,H>}
\\
\end{center}

For the experiment, the utilization range for the resources as mentioned in table \ref{tabrange} were followed. \textit{(Note - the limit $\infty$ is used to represent the maximum upper limit.)}

\begin{table}[H]
\centering
\scriptsize

\begin{tabular}{|c|c|c|c|}
\hline
\textbf{-} & \textbf{L} & \textbf{M} & \textbf{H} \\ \hline
CPU  & [0,20) & [20,70) & [70,100] \\ 
Usage(\%) & & & \\ \hline
Network & [0,10K) & [10K,800K) & [800k, $\infty$) \\ 
Usage(B/sec) & & & \\ \hline
Disk I/O  & [0,10K) & [10K,200K) & [200K, $\infty$)\\ 
Usage (B/sec) & & & \\ \hline
\end{tabular}
\caption{Range for the Components of Resource Usage Vectors}
\label{tabrange}
\end{table}

\begin{figure}[!h]
\centering
\includegraphics[scale=0.35]{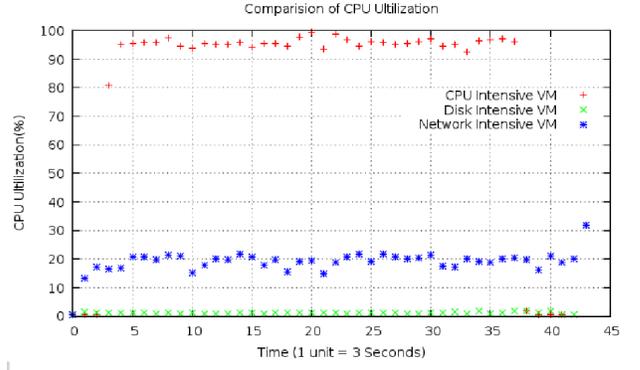}
\caption{Comparison of CPU Utilization}
\label{cpu_comparision}
\end{figure}

The CPU utilization of different VMs can be seen in Fig. \ref{cpu_comparision}. Average utilization of CPU Intensive VM remains around 80\%.The Average CPU utilization of Network Intensive VM, which is a web server, gives around 20\% CPU utilization, as it serves multiple request at a time, while that of a Disk intensive VM is zero. The lower limit for High CPU Utilization (H) is set at 70\% to incorporate any minor fluctuations that may occur, and bring down the average. For the lower limit of Medium CPU utilization, 20\% is used by taking the assumption that a Network Intensive VM will exhibit medium CPU utilization as in the current scenario, since not much server side processing is been done, the CPU utilization is bound to be above this level in any real scenario.

\begin{figure}[!h]
\centering
\includegraphics[scale=0.35]{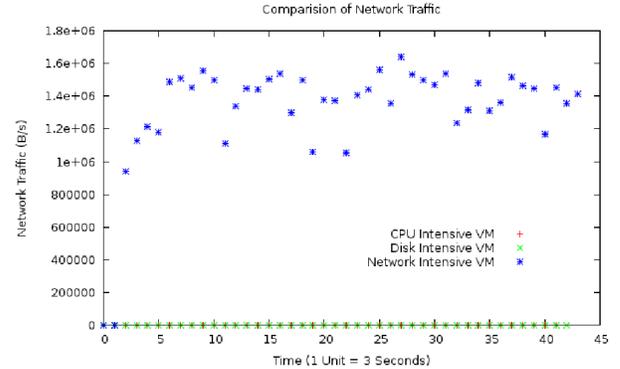}
\caption{Comparison of Network Traffic}
\label{network_comparision}
\end{figure}

The Network traffic by different VMs is given in Fig. \ref{network_comparision}. Here, for both CPU intensive VM, and for Disk Intensive VM, the Network traffic is almost zero, hence the lower limit for medium Network traffic is taken as 10KB/s to be on the safer side. The lower limit for High Network Traffic is taken at around 800KB/s.

\begin{figure}[H]
\centering
\includegraphics[scale=0.35]{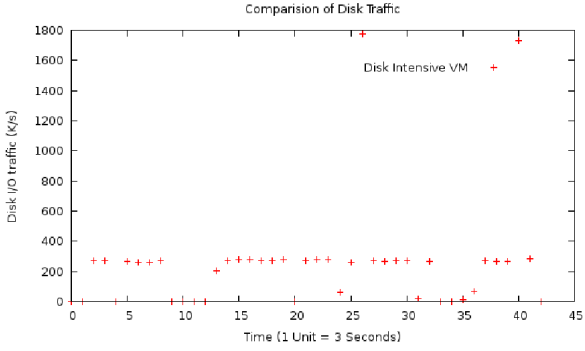}
\caption{Comparison of Disk Traffic}
\label{disk_comparision}
\end{figure}

The Disk traffic for CPU intensive and for the Network Intensive machines are almost zero, expect for a few fluctuations (not marked in Fig. \ref{disk_comparision}), therefore, considering the fact that Disk Intensive VM will exhibit high Disk traffic, the lower limit for High Disk traffic (H) is kept at 200 K/s (bytes). For lower limit of Medium Disk traffic, 10 K/s is used to accommodate any minor jitters (refer Fig. \ref{disk_comparision}).

The experiment gives an insight on the performance isolation provided by Xen, here it is clearly evident that resource contention results in performance drop, whether in the case of CPU intensive, Disk intensive, or Network intensive applications. It is important note that these resulting RUVs are generated after running benchmarks in stressed modes, however, real enterprise applications, like web servers, often run under normal mode, and hence the RUV may change accordingly. However, it is also clearly evident that placing a VM with another VM, which is using different type of resource, and hence having its vector \textit{orthogonal} to that of the former VM, results in mutual benefit for both.

In short, in order to have least amount of resource contention amongst VMs, the probability that two VMs which extensively use similar kind of resources be placed on the same host should be minimized. The authors exploit this \textit{Co-existential behaviour} in the following part of the paper. 

\textit{Note: The condition that the VMs on a particular physical server so not exceed the available resources is assumed implicit to the algorithm, and can be easily incorporated in the production environment}

\section{The VUPIC Algorithm} \label{secPA}
\subsection{Motivation}
The results obtained in experiment 1 show that VMs perform better in absence of resource contention. The RUV mentioned in equation \ref{eqRUV} provides an efficient method of quantifying the resource usage across different resources and can be employed in placement decisions. The proposed algorithm re-arranges the VMs according to their RUVs. The main objective of VUPIC is to club together those VMs which consume different resources, so as to minimize resource contention and also exploiting the resources available. VUPIC also ensures that no to VMs which extensively use same kind of resources are placed together, unless the situation can't be avoided. However, it will place two VMs using same resources if one of the VM uses the resource less extensively or normally.

{ \it The novel algorithm applies an optimization while making rearrangements on the basis of VMs corresponding RUVs. VUPIC would start arranging the VMs on PMs with a priority assignment that in the new arrangement, if possible, the VM should remain on the original host, in order to save VM migration costs. This will occur when the VUPIC is filling the PMs (bins) with the VMs and if the original or earlier PM is capable enough of holding the VM by satisfying the resource contention/ performance isolation requirements, the VM should remain on the original host so that one VM migration migration can be saved. This optimization might save many VM migration costs. This is applied as mentioned in line no. 11 in procedure 1. }

There might occur a case where a VM may not find any suitable host according to the proposed method, such a VM will be referred to as \textit{Compromised VM}. VUPIC provides a data structure ``compromisedVMList'' which stores all such compromised VMs. After allocation of all the VMs according to the algorithm, VUPIC places the \textit{Compromised VMs} in a manner such that the performance loss is minimized (refer Line no. 26 - 31 in procedure 1). For example, assume two physical host, M1 and M2, and three VMs V1, V2 and V2 with RUVs (H, M, L), (M, H, L), and (M, H, L). VUPIC will place V1 on M1, and V2 on M2. Now, according to the algorithm, V3 will be placed along with V2, as CPU has more priority than network, and since V2 is hosted on M2, V3 will be placed on M2.

The present algorithm can easily work with algorithms designed to manage the elasticity of the resources owned by the VMs.
After rearranging the VMs according to he proposed algorithm, the RUV may change, i.e., it may either increase or decrease, as the previous performance of machine could have been suppressed or compromised due to resource contention with similar VMs.
\subsection{Invocation of VUPIC}
The appropriate time of invocation of VUPIC depends on the user. One way would be to regularly call VUPIC after fixed duration of time and generate the placement schedule. Other way is to invoke the VUPIC when a client up-scales or down-scales its resource requirement and corresponding component of RUV changes. VUPIC can also be coupled with a resource manager which manages the scalability requirements of VMs on physical hosts, and in such case, VUPIC will be invoked if even after scaling the VM on the host, its resource vector does not change for a particular amount of time.

\begin{figure}[H]
\hrule
{\strut\footnotesize Procedure 1 Virtual Machine Usage Based Placement in Iaas Cloud (VUPIC)}
\hrule
\scriptsize
\begin{verbatim}

Input:  1)Set of RUVs <C,N,D> of each VM (MachinePool)
        2)present placement configuration of VMs
Output: 3)New placement configurations of VMs

00: BEGIN
01: Sort(MachinePool, DESC, vm.RUV)
02: compromisedVM = []
03: for all physicalHosts{
04:   set physicalHosts.state = < H, H, H >
06: }
07: while MachinePool is not Empty {
08:   placed = no
09:   VM = MachinePool.remove()
10:   Sort(PhysicalHosts, DESC, PhysicalHost.state)
11:   place VM's original host in front of the queue
12:   for all PM in physicalHosts{
13:     if (PM.state > VM.ruv){
14:       place(VM,PM)
15:       placed = yes
16:       break
17:     }
18:   }
19:   if (placed = no){
20:     for all PM in physicalHosts:
21:       if (PM.state == VM.ruv){
22:         place(VM,PM)
23:         placed = yes
24:         break
25:       }
26:     }
27:   if (placed = no){
28:     compromisedVM.insert(vm)
29:   }
30: }
31: while compromisedVMList is not Empty {
32:   vm = compromisedVMList.remove()   
33:   Sort(PhysicalHosts, DESC, PhysicalHost.state)
34:   place(vm,pm)
35: }
36:   
37: function place(vm,pm):
38:   for j = 1 to 3 : {   
39:     pm.state[j] = pm.state[j] - v.ruv.[j]    
40:   }
41:    
42: function Sort(List, order, value) 
43:  # Sorts the given list 
44:  # by the specified order and value
45: END

\end{verbatim}
\label{Algo1}
\end{figure}

\section{Experiment 2(a) : VUPIC}
\label{secE2}

\subsection{Experimental set-up}
\begin{figure}[t]
\centering
\subfigure[Initial Setup]{
\includegraphics[scale=0.4]{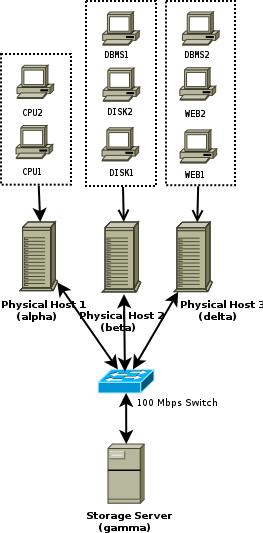}
\label{FinitPlac}
}
\subfigure[Rearranged Setup]{
\includegraphics[scale=0.4]{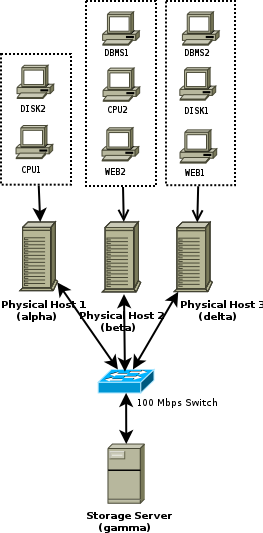}
\label{FnuPlac}
}
\label{exp2subfig}
\caption{Setup for Experiment 2(a)}
\end{figure}

The applications that were used in experiment 1 only intensively exploits the basic resources. The modern data centres and cloud host applications where the HPC and web server resemble the VMs used in experiment 1, hence to account for this property of cloud,  two VMs which run Database server are added to the cloud. The VM classification for experiment 2(a) is as specified in table \ref{TVM2}, and in figure \ref{FinitPlac}. The machines CPU1 and CPU2 are cloned instances of CPU intensive test VMs specified in section \ref{subsecTestvm}, similarly for DISK1 and DISK2, for WEB1 and WEB2, and for DBMS1 and DBMS2.

\begin{table}[t]
\centering
\begin{tabular}{|c||r|}
\hline
\textbf{Type} & \textbf{Machines}  \\ \hline
\textit{HPC} & \texttt{CPU1, CPU2} \\ \hline
\textit{Disk I/O} & \texttt{DISK1, DISK2}\\ \hline
\textit{HTTP Server} & \texttt{WEB1, WEB2} \\ \hline
\textit{DBMS server} & \texttt{DBMS1, DBMS2} \\ \hline
\end{tabular}
\caption{VM Classification for Experiment 2(a)}
\label{TVM2}
\end{table}

\subsection{Tests}
The experiment was conducted by executing the following benchmarks:
\begin{itemize}

\item \textit{CPU Intensive :} \texttt{sysbench} was run in CPU test mode for 120 seconds to calculate the prime numbers from 1 to 80,000. The result generated was the number of events of prime calculation that occurred in 120 seconds.

\item \textit{Disk Intensive :} \texttt{sysbench} was run in fileIO mode for 120 seconds to perform random read and write operations on 128 files of 2GB, the result generated here too was the number of events that occurred in 120 seconds.

\item \textit{Network Intensive :} the \textit{Apache HTTP} server was installed on the VM, serving a static HTML web page, and was benchmarked as follows, using \texttt{ab}, to serve 50,000 requests with 20 concurrent request in each iteration. The result generated here was the total time taken by the test, in seconds.

\item \textit{ DBMS server :} \texttt{sysbench} was run in OLTP mode for 120 sec, performing a series of write operations, the result generated was the number of successful transactions.
\end{itemize}

\subsection{The Resource Usage Vector (RUV)}
The resource usage vector was calculated by same method as mentioned previously in subsection \ref{SubDiscuss}, and the ranges for specifying L, M and H for each component of CPU, Network, and Disk are as specified in table \ref{tabrange}. The generated RUVs are mentioned in table \ref{Tab2}.

\begin{table}[H]
\centering
\scriptsize
\begin{tabular}{|l|c|c|c|}
\hline
\textit{\texttt{Machine}} & \textit{\texttt{CPU}} & \textit{\texttt{Network}} &  \textit{\texttt{Disk I/O}}\\
\hline
\texttt{CPU1 (CPU)} & \texttt{H} & \texttt{L} & \texttt{L}\\ \hline
\texttt{CPU2(CPU)} & \texttt{H} & \texttt{L} & \texttt{L}\\ \hline
\texttt{DISK1 (Disk I/O)} & \texttt{L} & \texttt{L} & \texttt{H} \\ \hline
\texttt{DISK2 (Disk I/O)} & \texttt{L} & \texttt{L} & \texttt{H} \\ \hline
\texttt{WEB1 (HTTPd)} & \texttt{M} & \texttt{H} & \texttt{L}\\ \hline
\texttt{WEB2 (HTTPd)} & \texttt{M} & \texttt{H} & \texttt{L}\\ \hline
\texttt{DBMS1 (MySQL)} & \texttt{L} & \texttt{L} & \texttt{H}\\ \hline
\texttt{DBMS2 (MySQL)} & \texttt{L} & \texttt{L} & \texttt{H}\\ \hline
\end{tabular}
\caption{Resource Usage Vectors for the VMs}
\label{Tab2}
\end{table}

The setup after considering the resource usage vectors (refer table \ref{Tab2}) and placing the machines according to the algorithm, the new rearrangement was generated. The change in the placement is shown in table \ref{plactab}, and in figure \ref{FnuPlac}.

\begin{table}[H]
\centering
\scriptsize
\begin{tabular}{|l|c|c|c|}
\hline
\textit{\texttt{Machine}} &\textit{\texttt{Initial Host}} &  \textit{\texttt{New Host}}\\
\hline
\texttt{CPU1} & \texttt{alpha} & \texttt{alpha}\\ \hline
\texttt{CPU2} & \texttt{alpha} & \texttt{beta}\\ \hline
\texttt{DISK1} & \texttt{beta} & \texttt{delta} \\ \hline
\texttt{DISK2} & \texttt{beta} & \texttt{alpha} \\ \hline
\texttt{WEB1} & \texttt{delta} & \texttt{delta}\\ \hline
\texttt{WEB2} & \texttt{delta} & \texttt{beta}\\ \hline
\texttt{DBMS1}& \texttt{beta} & \texttt{beta}\\ \hline
\texttt{DBMS2}& \texttt{delta} & \texttt{delta}\\ \hline
\end{tabular}
\caption{Hosts for the VMs}
\label{plactab}
\end{table}

\subsection{Results of Experiment 2(a)}

\begin{table}[H]
\centering
\scriptsize
\begin{tabular}{|l|c|c|}
\hline
\textit{\texttt{Machine}} & \textit{\texttt{Performance (before)}} & \textit{\texttt{Performance (after)}} \\
\hline
\texttt{CPU1} & \texttt{4286\#} & \texttt{5527\#} \\ \hline
\texttt{CPU2} & \texttt{4812\#} & \texttt{5199\#} \\ \hline
\texttt{DISK1} & \texttt{1466\#} & \texttt{2830\#} \\ \hline
\texttt{DISK2} & \texttt{2200\#} & \texttt{5600\#} \\ \hline
\texttt{WEB1} & \texttt{136s} & \texttt{71.367s} \\ \hline
\texttt{WEB2} & \texttt{136s} & \texttt{94s} \\ \hline
\texttt{DBMS1} & \texttt{208\#} & \texttt{725\#} \\ \hline
\texttt{DBMS2} & \texttt{686\#} & \texttt{365\#} \\ \hline
\end{tabular}
\caption{Comparison of Results Obtained in Experiment 2(a) Before and After Re-arrangement of VMs}
\label{Tab4}
\end{table}
All the performances are given in terms of number of events completed by the test, except for web servers, where it stands for time taken to complete the test (in seconds).

The comparison of the results obtained is provided in table \ref{Tab4}. Here, the mutual benefit that is provided to VMs after rearranging them according to the proposed algorithm is very promising. For the case of the VMs CPU1 and CPU2, while their performance before applying the algorithm was 4286 and 4812 events respectively, upon relocating CPU1 with DISK2, the performance of CPU1 rises to 5527 events, for CPU2 (relocated with DBMS1 and WEB2), the performance rises to 5199 events.

Similarly, if we observe the machines originally placed on physical host - \textit{beta}, the performance of DISK1 rises form 1466 to 2830 events when placed along with DBMS2 and WEB1, so does for  DISK2, where it performance rises from 2200 events to 5600 events on placing with CPU1, this more than twofold increase in case of DISK2 can be attributed to the absence of disk intensive VMs on physical host \textit{alpha}, which is not the case for DISK1, where one disk intensive VM is already present (DBMS2 on \textit{delta}). For the case of VM DBMS1, its performance also increases as from 208 to 725 transaction (more than three fold increase), upon relocation of DISK1 and DISK2 from \textit{beta}.

For the machines on physical host delta, the performance of VM WEB1 becomes double, as its response time is reduced to almost half (from 136 seconds to ~71 seconds) when placed with DISK1 and DBMS2, similarly for WEB2, the performance in increased from 136 seconds to 94 seconds, upon its relocation on physical host beta, with DBMS1 and CPU2. The case of compromised placement is also visible here, where the performance of DBMS2 decreases to its half, i.e., drops from 686 events to 365 events, and the difference in the increase that DISK1 and DISK2 experience (increase by 1364 events in case of DISK1, while increase by 3400 events for DISK2) can be seen as the side-effect caused by placing the compromised VM DBMS2 on machine \textit{delta}, which also hosts DISK1 (both DISK1 and DBMS2 share extensively use Disk I/O, hence resource contention comes into picture again).

\section{Experiment 2(b) - VUPIC with multiple VCPU considerations}
In real world, most of the applications are multi-threaded, and also may have more than one VCPU. It is recommended to keep the number of VCPUs equal to that of PCPUs \cite{cpuover}, experiment 2(b) is presented to show the capability of the algorithm to take the fact into account  that a VM may have more than one VCPUs.
Most of the applications in cloud are multi-threaded, like web server and DBMS, A modified version of the proposed algorithm is given in figure~\ref{Algo2} which will also factor the number of VCPUs allotted to a VM. For any physical host, there cant be more VCPUs present on it than the number of PCPUs available, and the same is formulated in subsection \ref{formu}.
\subsection{Upper Limit on VCPUs} \label{formu}
In this version of algorithm, zero CPU overcommitment policy is followed as it is recommended by Xen enterprise server vendors \cite{cpuover}. Therefore the mathematical formulation for this case will be,
{\scriptsize
$$ \nonumber \forall j\;\; Such\;That\;\;\; PM_{j}\;is\;a\;Physical\;Server $$
\begin{equation}
\sum\limits_{ VM_{i} \epsilon VM\;on\;PM_{j}} VCPU_{vm_{i}} \leq \;number\;of\; PCPUs\; on\; PM_{J}
\end{equation}
}
However, the overcommitment policy can be changed by simple increasing the PCPU limit of a physical host.
\subsection{Experimental setup}
\begin{figure}[H]
\centering
\subfigure[Initial Setup]{
\includegraphics[scale=0.42]{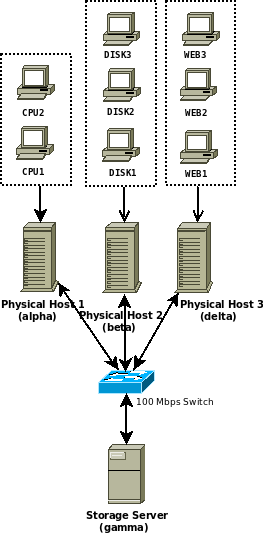}
\label{3FinitPlac}
}
\subfigure[Rearranged Setup]{
\includegraphics[scale=0.42]{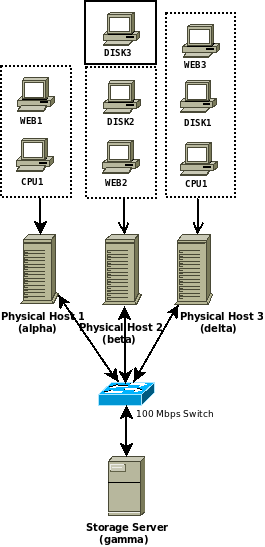}
\label{3FnuPlac}
}
\label{exp3subfig}
\caption{Setup for Experiment 2(b)}
\end{figure}
The initial VM arrangement is same as that in experiment 2(a) (refer figure \ref{3FinitPlac}), with the role of DBMS1 changed to that of a disk intensive VM. It will be called DISK3. Similarly, the role of DBMS2 changed to a web server (named WEB3), and each VM may have one or two VCPUs (refer table \ref{plac3tab}). The benchmarks used are same as mentioned in section \ref{subsecTestvm}), with the exception of CPU2, DISK2, DBMS1, WEB2, DBMS2, where each will use two threads (refer table \ref{plac3tab}).

\begin{figure}[H]
\hrule
{\strut\footnotesize Procedure2: VUPIC (with CPU overcommitment constraints)}
\hrule
\scriptsize
\begin{verbatim}
Input:  1)Set of RUVs <C,N,D> of each VM along with 
          their VCPU requirement (MachinePool)
        2)present placement configuration of VMs
Output: 1)New placement configurations of VMs
00: BEGIN
01: Sort(MachinePool, DESC, vm.RUV)
02: compromisedVM = []
03: for all physicalHosts{
04:   set physicalHosts.state = < H, H, H >,
05:   physicalHosts.PCPU = # number of processors available
06: }
07: while MachinePool is not Empty {
08:   placed = no
09:   VM = MachinePool.remove()
10:   Sort(PhysicalHosts, DESC, PhysicalHost.state)
11:   place VM's original host in front of the queue
12:   for all PM in physicalHosts{
13:     if (PM.state > VM.ruv) and (PM.pcpu >= VM.vcpu){
14:       place(VM,PM)
15:       placed = yes
16:       break            }}
17:     if (placed = no){
18:       for all PM in physicalHosts:
19:       if (PM.state == VM.ruv) and (PM.pcpu >= vm.vcpu){
20:         place(VM,PM)
21:         placed = yes
22:         break            }}
23:     if (placed = no){
24:         compromisedVM.insert(vm) }}
25: while compromisedVMList is not Empty {
26:   vm = compromisedVMList.remove()   
27:   Sort(PhysicalHosts, DESC, PhysicalHost.state)
28:   placed = no
29:   for all pm in PhysicalHosts{
30:     if (pm.pcpu >= vm.vcpu){
31:       place(vm,pm)
32:       placed = yes     }}
33:   if (placed == no)
34:       print "VM ",vm, "INVALID/UNFIT" 
35: }
35: function place(vm,pm):
36:   for j = 1 to 3 : {   
37:     pm.state[j] = pm.state[j] - v.ruv.[j]    
38:   }
39:   pm.pcpu = pm.pcpu - vm.vcpu
40: function Sort(List, order, value) 
41:  # Sorts the given list 
42:  # by the specified order and value    
43: END
\end{verbatim}
\label{Algo2}
\caption{VUPIC (with CPU overcommitment constraints)}
\end{figure}
\begin{table}[H]
\centering
\scriptsize
\begin{tabular}{|l|c|c|c|}
\hline
\textit{\texttt{Machine}} &\textit{\texttt{Initial Host}} & \textit{VCPUs and Threads}\\
\hline
\texttt{CPU1} & \texttt{alpha} & \texttt{1}\\ \hline
\texttt{CPU2} & \texttt{alpha} & \texttt{2}\\ \hline
\texttt{DISK1} & \texttt{beta} & \texttt{1} \\ \hline
\texttt{DISK2} & \texttt{beta} & \texttt{2} \\ \hline
\texttt{DISK3}& \texttt{beta} & \texttt{2}\\ \hline
\texttt{WEB1} & \texttt{delta} & \texttt{1}\\ \hline
\texttt{WEB2} & \texttt{delta} & \texttt{2}\\ \hline
\texttt{WEB3}& \texttt{delta} & \texttt{2}\\ \hline
\end{tabular}
\caption{Hosts for the VMs and VCPUs Allotted}
\label{plac3tab}
\end{table}

\subsection{Results of Experiment 2(b) and Discussion}
The generated migration schedule is (table \ref{MigSch3tab}):
\begin{table}[H]
\centering
\scriptsize
\begin{tabular}{|l|c|c|c|}
\hline
\textit{\texttt{Machine}} &\textit{\texttt{Initial Host}} & \textit{New Host}\\
\hline
\texttt{CPU1} & \texttt{alpha} & \texttt{delta}\\ \hline
\texttt{CPU2} & \texttt{alpha} & \texttt{alpha}\\ \hline
\texttt{DISK1} & \texttt{beta} & \texttt{delta} \\ \hline
\texttt{DISK2} & \texttt{beta} & \texttt{beta} \\ \hline
\texttt{DISK3}& \texttt{beta} & \texttt{INVALID/UNFIT}\\ \hline
\texttt{WEB1} & \texttt{delta} & \texttt{alpha}\\ \hline
\texttt{WEB2} & \texttt{delta} & \texttt{beta}\\ \hline
\texttt{WEB3}& \texttt{delta} & \texttt{delta}\\ \hline
\end{tabular}
\caption{Hosts for the VMs and VCPUs Allotted}
\label{MigSch3tab}
\end{table}

After running the experiment as mentioned above, the VM DISK3 was not assigned to any physical host (refer figure \ref{3FnuPlac}),  as it requires 2 VCPUs and assigning it to any physical server would breach the PCPU limit. In such a case VUPIC declares such a VM to be INVALID/UNFIT for placement. This behaviour is enforced so that the other VMs may not be affected by the high need of VCPUs by DISK3, however, it can be placed with a high degree of compromise, or can be migrated to another cloud, and this decision can be made by the user and can be configured in VUPIC.

\section{Implementation of VUPIC}
\label{secIoA}
In current version, VUPIC is a collection of interdependent python scripts, which handles the task of logging the resource usage done by the VMs, producing  RUVs using the mean resource usage from the generated log, and generating placement schedule. The VUPIC clients generate the resource usage logs and using them generate the RUVs as mentioned in equation \ref{eqRUV}, and the ranges for the components used are as specified in table \ref{tabrange}. The resource generation script has a CPU overhead of around 3\%, and can be attributed to its collection of information from three different sources of xentop \cite{xentop}, ifstat \cite{ifstat}, and iotop \cite{iotop}. These ranges can be changed with ease, and as mentioned before, the statistical function used to generate the RUV can be replaced by any other function.
\begin{figure}[H]
\centering
\includegraphics[scale=0.3]{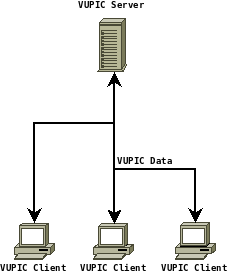}
\caption{VUPIC Architecture}
\label{figVUPICarch}
\end{figure}
The current version of the VUPIC client generates the RUVs and stores them on the storage server for each client as a file. The VUPIC server uses these files to generate the placement schedule. This schedule can be used by any migration manager to execute the migration.

\section{Related Work}
\label{secRW}
The problem of resource allocation is similar to that in Grids and Clusters. Some of the popular resource allocation mangers used in various environment are GRAM \cite{gram}, and Condor \cite{condor}. Some popular cloud management system are Open Nebula \cite{opennebula}, Eucalyptus \cite{eucalyptus} and Nimbus \cite{nimbus}. OpenNebula provides for \textit{Match Making Scheduler} where it does \textit{rank scheduling} as specified by the user. Resource allocation issue also arises when a VMs upscale their available resource to meet the increased demand.
Virtual machine placement is a critical and central issue related to resource allocation in cloud and it has been studied extensively. C. Hyser et. al. \cite{chyser} provides with a framework for autonomic placement manager. There are a number of virtual machine placement algorithms which aim to provide energy efficient scheduling, A. Verma et. al. \cite{averma} have studied the problem of power consumption in cloud and have proposed pMapper in \cite{pmapper}. B. Li et. al. proposed enaCloud in \cite{enacloud} which aims to achieve similar objective, another such research is done by J. Xu et. al. in \cite{jxu} which along with power, also takes into account thermal dissipation costs. Similar work is done by A. Beloglazov et. al. in \cite{anton}, and in \cite{kle}.
When looking at the VM placement problem from the point of view of maximizing the available physical resource usage, performance isolation becomes another critical issue that needs to be taken care of. J. N. Matthews et. al. provide a design for performance isolation benchmark in \cite{jeanna}. The performance isolation provided by Xen and the effect of I/O intensive applications with low latency has been studied in \cite{gsomani}. As the performance isolation can be mapped to low level of physical machine, the CPU scheduler also plays a role in deciding performance of virtual machines  as studied in \cite{dguptaCPUscheduler}. There are a number of resource allocation strategies which have been created from the point of view of maximizing physical resource utilization as studied in \cite{ubellur}, the problem reduces to that of the classical NP-hard problem of bin packing. Another approach is to reduce the network traffic due to intra-VM communications as studied in \cite{sujesha}. Network aware placement has been studied by J. T. Piao et. al. in \cite{jtpiao}.
These algorithms work with optimal migration algorithm, which may take into account the post-migration network load \cite{vivek} and minimizing the VM down time \cite{cclark}. The proposed novel VM placement algorithm significantly improves the VM performance,   which otherwise was compromised. The paper also introduces the concept of resource usage variation and provides efficient method to incorporate the same using Resource Usage Vectors (RUV) in placement decision. The algorithm generates a placement schedule, which can be easily deployed using any of the above mentioned migration techniques.

\section{Conclusion and Future Work}
\label{secCFW}
\label{conclusion}
The present work incorporates performance isolation and resource contention properties into account while taking VM placement as well as resource allocation decisions. Any infrastructure cloud would be a multi-tenant service provider and would host virtual machines of varied types on different physical servers. Virtual machines generate different resource usage and thus create a behavioral usage pattern. Experiments were conducted to show how individual VMs affect and get affected by the neighboring VMs on the same physical server. Currently, there is no virtual machine placement algorithm that takes the resource usage patterns into account.

The proposed novel algorithm takes resource usage by the virtual machines into account while making placement decisions, and also provides an efficient way to incorporate these resource usage patterns into the algorithms using a 3-dimensional vector called Resource Usage Vector (RUV). A modified version of the same algorithm has been presented which takes care of VMs with multiple VCPUs and also restricts the CPU over commitment. Both of these algorithms generate a migration schedule which minimizes the number of unnecessary migrations by allocating a VM to its original host if all the constraints are satisfied. The generated migration schedule can be executed using an optimal migration algorithm to reduce the migration costs.  Authors also plan to incorporate power efficiency and migration algorithms to implement it as a complete IaaS solution

%

\vfill
\begin{IEEEbiographynophoto}{Gaurav Somani}
is a faculty member at The LNM Institute of Information Technology, Jaipur, India. His research interest   include cloud computing, virtualization, ad-hoc networks and distributed computing. He has published in many reputed conferences like IEEE CLOUD 2009, IC3 2010, IEEE PDGC 2010. A monograph has been published on his recent works by VDM publishers on scheduling and isolation in virtualisation.
\end{IEEEbiographynophoto}%

\begin{IEEEbiographynophoto}{Prateek Khandelwal}
is a student at The LNM Institute of Information Technology, Jaipur, India.
\end{IEEEbiographynophoto}%


\begin{IEEEbiographynophoto}{Kapil Phatnani}
is a student at The LNM Institute of Information Technology, Jaipur, India.
\end{IEEEbiographynophoto}%




\end{document}